\documentclass[longauth]{aa}             
\usepackage{graphicx}
\usepackage{txfonts}
\def\kms{\hbox{\,km\,s$^{-1}$}}       

\def\degr{\hbox{$^\circ$}}

\begin{document}
   \title{Transiting exoplanets from the CoRoT space mission
   \subtitle{XXI. CoRoT-19b: A low density planet orbiting an old inactive F9V-star}
\thanks{The CoRoT space mission, launched on December 27, 2006, has
  been developed and is operated by the CNES, with the contribution of
  Austria, Belgium, Brazil, ESA (RSSD and Science Program), Germany
  and Spain. Partly based on observations obtained at the European Southern
  Observatory at Paranal, Chile in program 184.C-0639, and partly based
  on observations conducted at McDonald Observatory.}}
\authorrunning{Guenther et al.}
\titlerunning{CoRoT-19b}
\author{Guenther, E.W.\inst{1}, 
D\'{\i}az, R.F.\inst{2,5}, 
Gazzano, J.-C.\inst{6},  
Mazeh, T.\inst{4}, 
Rouan, D.\inst{9}, 
Gibson, N.\inst{7}, 
Csizmadia, Sz.\inst{11},         
Aigrain, S.\inst{7}, 
Alonso, R.\inst{8}, 
Almenara J.M.\inst{16},  
Auvergne, M.\inst{9},   
Baglin, A.\inst{9}, 
Barge, P.\inst{3}, 
Bonomo, A. S.\inst{3}, 
Bord\'e, P.\inst{10}, 
Bouchy, F.\inst{2,5},  
Bruntt, H.\inst{23},  
Cabrera, J.\inst{11}, 
Carone, L.\inst{14}, 
Carpano, S.\inst{15}, 
Cavarroc, C.\inst{10},
Deeg, H. J.\inst{16,21},
Deleuil, M.\inst{3}, 
Dreizler, S.\inst{22}, 
Dvorak, R.\inst{17}, 
Erikson, A.\inst{11}, 
Ferraz-Mello, S.\inst{18},
Fridlund, M.\inst{15}, 
Gandolfi, D.\inst{15},
Gillon, M.\inst{19}, 
Guillot, T.\inst{6}, 
Hatzes, A.\inst{1}, 
Havel, M.\inst{6}, 
H\'ebrard, G.\inst{2,5}, 
Jehin, E.\inst{19},
Jorda, L.\inst{3},
Lammer, H.\inst{20}, 
L\'eger, A.\inst{10}, 
Moutou, C.\inst{3}, 
Nortmann, L.\inst{22},  
Ollivier, M.\inst{10}, 
Ofir, A.\inst{4}, 
Pasternacki, Th. \inst{11},  
P\"atzold, M.\inst{14},
Parviainen, H. \inst{16,21}, 
Queloz, D.\inst{8}, 
Rauer, H.\inst{11,12}, 
Samuel, B.\inst{10}, 
Santerne, A.\inst{3,5}, 
Schneider, J.\inst{13}, 
Tal-Or, L.\inst{4}, 
Tingley, B.\inst{16,21}, 
Weingrill, J. \inst{20},
 Wuchterl, G. \inst{1} 
          }
   \institute{Th\"uringer Landessternwarte Tautenburg, 07778 Tautenburg, Germany
              (\email{guenther@tls-tautenburg.de})
                \and
Institut d'Astrophysique de Paris, UPMC-CNRS, UMR7095, Institut
d'Astrophysique de Paris, 75014, Paris, France
         \and
Laboratoire d'Astrophysique de Marseille, 38 rue Fr\'ed\'eric  Joliot-Curie, 
13388 Marseille Cedex 13, France 
         \and
School of Physics and Astronomy, Raymond and Beverly Sackler Faculty
of Exact Sciences, Tel Aviv University, Tel Aviv, Israel
         \and
Observatoire de Haute-Provence, Universit\'e Aix-Marseille \& CNRS, 04870
St. Michel l'Observatoire, France
         \and
Observatoire de la C\^ote d' Azur, Laboratoire Cassiop\'ee, BP 4229,
06304 Nice Cedex 4, France
         \and
Department of Physics, Denys Wilkinson Building Keble Road, Oxford, OX1 3RH, UK 
         \and
Observatoire de l'Universit\'e de Gen\`eve, 51 chemin des Maillettes, 1290
Sauverny, Switzerland
         \and
LESIA, UMR 8109 CNRS, Observatoire de Paris, UVSQ, Universit\'e
Paris-Diderot, 5 place J. Janssen, 92195 Meudon, France
         \and
Institut d'Astrophysique Spatiale, Universit\'e Paris-Sud 11, 91405 Orsay, France
         \and
Institute of Planetary Research, German Aerospace Centre,
Rutherfordstra\ss e 2, 12489 Berlin, Germany
         \and
Centre for Astronomy and Astrophysics, TU Berlin, Hardenbergstra\ss e 36,
10623 Berlin, Germany
         \and
LUTH, Observatoire de Paris, CNRS, Universit\'e Paris Diderot; 5 place
Jules Janssen, 92195 Meudon, France
         \and
Rheinisches Institut f\"ur Umweltforschung an der Universi\"at zu K\"oln,
Aachener Stra\ss e 209, 50931 K\"oln, Germany
         \and
Research and Scientific Support Department, ESTEC/ESA, PO Box 299, 2200
AG Noordwijk, The Netherlands
         \and
Instituto de Astrof\'\i sica de Canarias, 38205 La Laguna, Tenerife, Spain
         \and
University of Vienna, Institute of Astronomy, T\"urkenschanzstra\ss e 17,
         1180 Vienna, Austria
         \and
Instituto de Astronomia, Geof\'\i sica e Ci\^encias Atmosf\'ericas,
Universidade de S\~ao Paulo, Rua do Mat\~ao 1226, 05508-900 S\~ao
Paulo, Brazil 
         \and
University of Li\` ege, All\'ee du 6 ao\^ut 17, S. Tilman, Li\` ege 1, Belgium
         \and
Space Research Institute, Austrian Academy of Science, Schmiedlstr. 6,
8042 Graz, Austria
         \and
Dpto. de Astrof\'\i sica, Universidad de La Laguna, 38206 La Laguna,
Tenerife, Spain
         \and
Georg-August-Universit\"at, Institut f\"ur Astrophysik, 
Friedrich-Hund-Platz 1,  37077  G\"ottingen, Germany
         \and
Department of Physics and Astronomy, Aarhus University, 8000 Aarhus C, Denmark
}

\date{Received 15, 2011; accepted Oct 20, 2011}
  \abstract 
{Observations of transiting extrasolar planets are of key importance
  to our understanding of planets because their mass, radius, and 
  mass density can be determined. These measurements indicate that 
  planets of similar mass can have very different radii. For 
  low-density planets, it is generally assumed that they are inflated 
  owing to their proximity to the host-star. To determine the causes
  of this inflation, it is necessary to obtain a statistically significant
  sample of planets with precisely measured masses and radii.}
{The CoRoT space mission allows us to achieve a very high photometric 
  accuracy. By combining CoRoT data with high-precision radial velocity 
  measurements, we derive precise  planetary radii and masses. We report
  the discovery of CoRoT-19b, a gas-giant planet transiting an old, inactive 
  F9V-type star with a period of four days.}
{After excluding alternative physical configurations mimicking a
  planetary transit signal, we determine the radius and mass of the
  planet by combining CoRoT photometry with
  high-resolution spectroscopy obtained with the echelle spectrographs
  SOPHIE, HARPS, FIES, and SANDIFORD. To improve the precision
  of its ephemeris and the epoch, we observed additional transits with
  the TRAPPIST and Euler telescopes. Using HARPS spectra obtained during
  the transit, we then determine the projected angle between the spin
  of the star and the orbit of the planet.}
{We find that the host star of CoRoT-19b is an inactive F9V-type star
  close to the end of its main-sequence life. The host star has a mass
  $M_{*}=1.21\pm0.05$ \,$M_\odot$ and radius
  $R_{*}=1.65\pm0.04$\,$R_\odot$.  The planet has a mass of
  $M_P=1.11\pm0.06$\,$M_{Jup}$ and radius of
  $R_P=1.29\pm0.03$\,$R_{Jup}$. The resulting bulk density is
  only $\rho=0.71\pm0.06$\,$g\,cm^{-3}$, which is much lower than 
  that for Jupiter.}
{The exoplanet CoRoT-19b is an example of a giant planet of almost 
  the same mass as Jupiter but a $\approx $30\% larger radius. }
   \keywords{planetary systems --
             techniques: photometric --
             techniques: radial velocities --
             techniques: spectroscopic --
              }
   \maketitle
%

\section{Introduction}

The discovery of gas-giant planets orbiting their star at distances
$\leq0.1$\,AU more than 15 years ago was surprising, and their
study is still an active field of research.  A key observation for
understanding these objects are those of transiting planets because 
they allow us to determine their true mass, radius, and full orbital parameters. 
For some transiting planets, it is even possible  to determine
the projected angle between the stellar rotation and orbital spin axes. Thus,
transiting planets allow us to obtain very detailed information that
is essential to constrain models of  the formation and
evolution of extrasolar planets.

The determination of the mass and radius and hence the density of
extrasolar planets led to a rather unexpected result that planets of
the same mass can have very different radii and thus very different
bulk densities. Planets of very low density are possibly inflated owing
to their close proximity to the host star. To aid our
understanding of why close-in planets are inflated, it is essential
to obtain precise values for the masses and radii of many extrasolar
planets orbiting stars of different mass, temperature, luminosity, and
orbital distances.  Since accurate planetary parameters are crucial
for modeling the structure of the planet, it is preferable to have
precise measurements of a small sample of transiting exoplanets,
rather than low precision measurements of a larger sample. Space-based
transit search programs are ideal for this purpose because their high
sensitivity and excellent time-coverage enables us to explore a wider
range of planet densities, and orbital periods.

Here we present the discovery of \object{CoRoT-19b}, a planet with
the mass of Jupiter but a radius that is substantially larger. By
determining the properties of the planet and its host star precisely,
we thus add in an important piece to the puzzle of exoplanets.

\section{The CoRoT light-curves}

\begin{figure}
 \includegraphics[height=.30\textheight,angle=0.0]{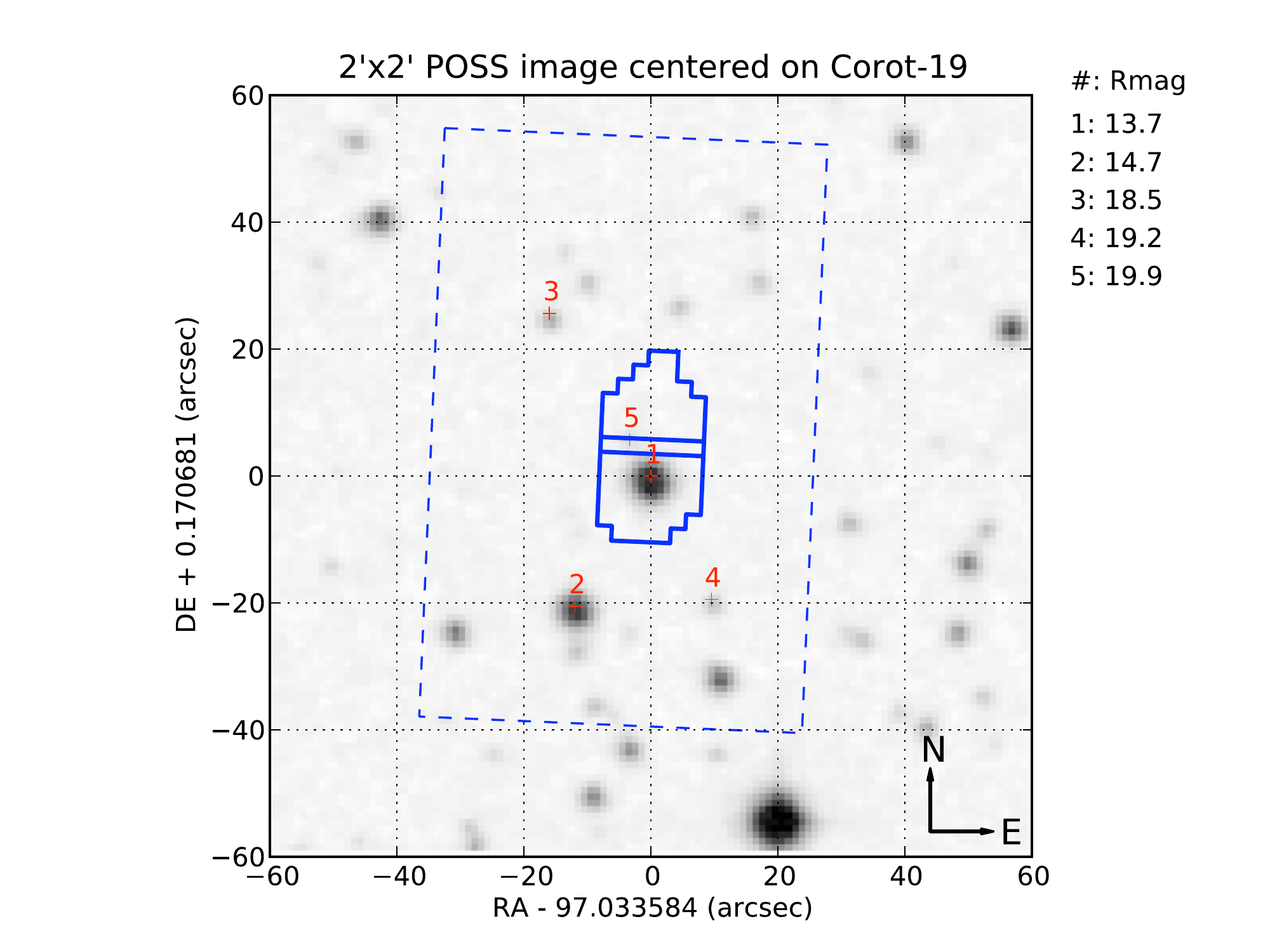}
 \caption{DSS image of the field. The jagged outline at the centre is
   the photometric aperture mask (North is up, east is right).}
  \label{mask}
 \end{figure}

\object{CoRoT-19} was continuously observed with the CoRoT satellite
from March 5 to 29, 2010 in the chromatic mode. The coordinates,
proper motion, and brightness of the star are given in Table~\ref{tab:star}.  
Fig.\,\ref{mask} shows an DSS image of the field with the mask image 
superimposed.

\begin{figure}
  \includegraphics[height=.25\textheight,angle=0]{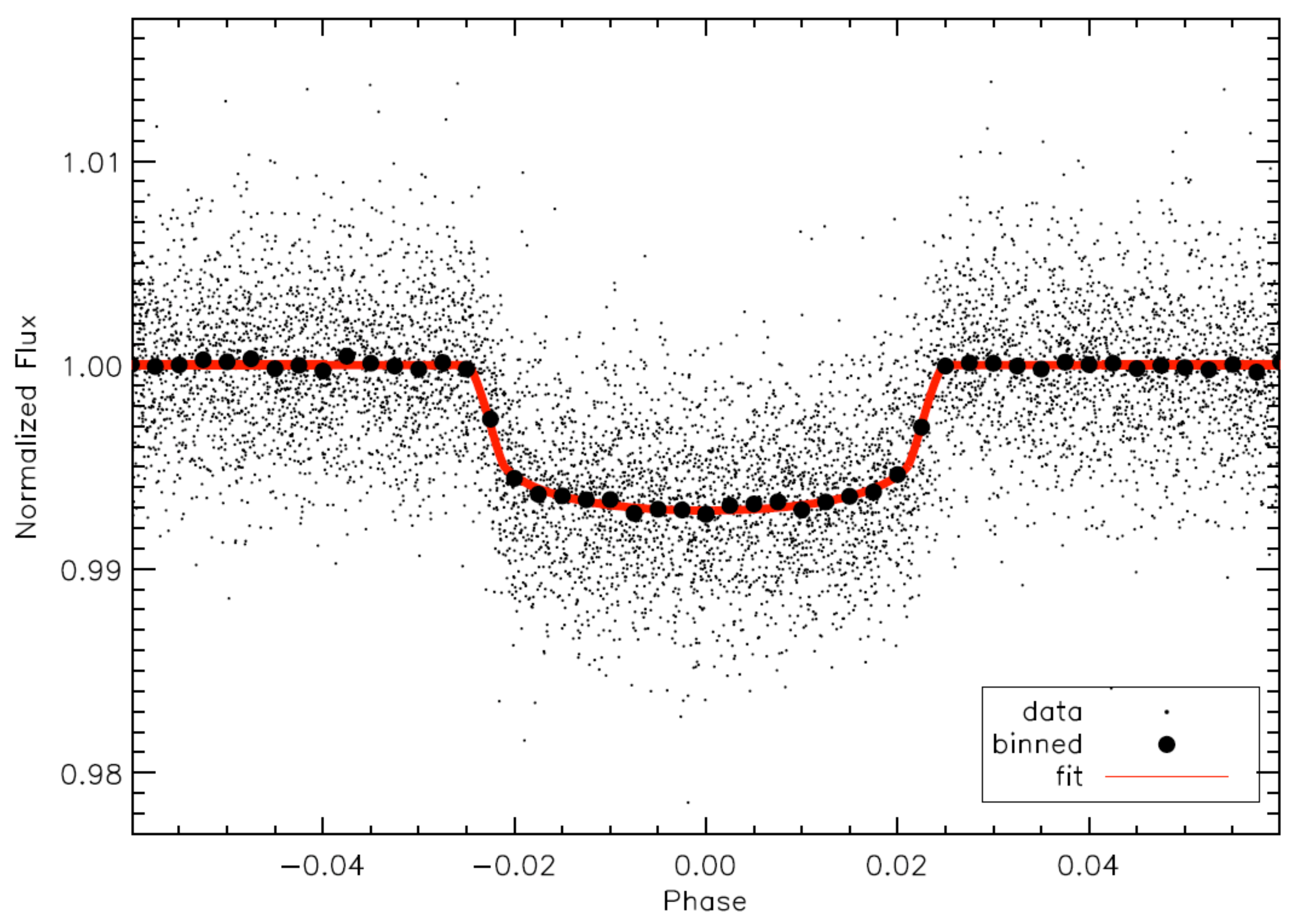}
\caption{Light-curve of CoRoT-19 obtained by CoRoT in white-light together
         with the best-fit model.}
  \label{LC}
\end{figure}

The sampling of the light-curve was 32\,seconds, thus 65123
photometric measurements in three colours were obtained.  
Six full transit-like events were visible, and a seventh one was partly
covered. The transit-like events are periodic with a period of
$3.8971272\pm0.000020$ days. The depth of successive transits was
constant to within $0.79\pm0.02$\,\%.  Fig.\,\ref{LC} shows the
monochromatic light-curve of the transit produced by the standard
CoRoT pipeline (version 2.1), after combining the signal from the 
three-colour channels (Auvergne et al. \cite{auvergne09}) and 
co-adding all transits.

\begin{table}
 \caption{CoRoT-19, coordinates, and brightness magnitudes}
\setlength{\tabcolsep}{3pt} 
 \begin{tabular}{cc}
\noalign{\smallskip}
 \hline 
\noalign{\smallskip}
CoRoT window ID    & SRa03\,E2\,0490 \\
CoRoT ID           & 315198039 \\
RA (2000)          & $06^{h} 28^{m} 08\fs{04}^1$ \\                   
Dec (2000)         & $-00\degr 10\arcmin {14}\farcs{4}^1$ \\
B                  & $14.780^1$ \\ 
V                  & $14.007^4$  \\ 
R                  & $13.482\pm0.065^5$ \\ 
I                  & $12.996\pm0.03^2$ \\
J                  & $12.324\pm0.024^3$ \\
H                  & $11.961\pm0.022^3$ \\ 
K                  & $11.837\pm0.024^3$ \\ 
\noalign{\smallskip}
 \hline 
 \end{tabular}
 \label{tab:star}
\\
$^1$ USNO-B1/B2,
$^2$ DENIS ,
$^3$ 2MASS,
$^4$ CALC-2MASS,
$^5$ CMC 14 
\end{table}

The fitting of the transit lightcurve was carried out by two groups
independently and both obtained consistent results
(Table~\ref{tab:planet}).  For the light-curve modeling, we used a
contamination factor of $\leq 0.32\%$ and a circular orbit (see
Sections~\ref{Sec:onoff} and \ref{Sec:RV}).

The first analysis used the Markov chain Monte Carlo (MCMC) algorithm
described by Tegmark et al. (\cite{tegmark04}) and Holman et
  al. (\cite{holman06}) for the implementation described in Gibson
  et al. (\cite{gibson10}). Using this method, we found that
$a/R_{*}=6.79\pm{0.20}$, $R_p/R_{*}=0.07699^{+0.0008}_{-0.0007}$, and
$(M_{*}/M_{\odot})^{1/3}(R_*/R_{\odot})^{-1}=0.65^{+0.01}_{-0.03}$.

\begin{table}
  \caption{Transit times}
\setlength{\tabcolsep}{3pt} 
 \begin{tabular}{llrl}
\noalign{\smallskip}
 \hline 
\noalign{\smallskip}
  Instrument & Mid-transit time & O-C & transit \\
  used       & HJD [UTC]           & [min] & No. \\
\noalign{\smallskip}
 \hline  
\noalign{\smallskip}
CoRoT    & $2455 261.3376\pm0.0013$ & -0.78 & 1 \\
CoRoT    & $2455 265.2383\pm0.0017$ & 4.36 & 2\\ 
CoRoT    & $2455 269.1320\pm0.0016$ & -0.57 & 3 \\ 
CoRoT    & $2455 273.0303\pm0.0012$ & 1.12 & 4 \\ 
CoRoT    & $2455 276.9258\pm0.0014$ & -1.23 & 5 \\ 
CoRoT    & $2455 280.8204\pm0.0016$ & -4.86 & 6 \\ 
CoRoT    & $2455 284.7218\pm0.0021$ & 1.29 & 7\\ 
TRAPPIST & $2455 627.6706\pm0.0022$ & 3.60 & 95 \\
{\it Euler}    & $2455 635.4666\pm0.004$ & 6.12 & 97 \\
\noalign{\smallskip}
 \hline 
\noalign{\smallskip}
 \end{tabular}
 \label{tab:transit}
\end{table}

Independently of the MCMC method, we carried out another analysis
using a genetic algorithm optimization. This allowed us to verify the
uniqueness of the solution. The data was fitted using the model of
Mandel \& Agol (\cite{mandel02}) after it had been phase-folded.  The free
parameters are: the epoch, the ratio of the semi-major axis ($a$) to
the star's radius ($R_\mathrm{*}$), the ratio of the radius of the
planet to stellar radius $k=R_\mathrm{p}/R_\mathrm{*}$, and the impact
parameter ($b=a\cos i/R_\mathrm{*}$, with the inclination $i$).  We
used the limb darkening coefficients published by Sing (\cite{sing01})
as defined by Brown et al. (\cite{brown01}) using the $T_{\rm
  eff}$, $\log\,g$-values, and metallicity of the star derived in
Section~\ref{Sec:SpectralAnalysis} and given in
Table~\ref{tab:planet}. The accuracy is thus limited only by the
accuracy with which the stellar parameters were determined.  The
phase-folded light-curve of the transit together with the best-fit
solution obtained is shown in Fig.\,\ref{LC}.  The results and the
1$\sigma$ uncertainties are $a/R_{*}=6.68\pm0.13$,
$R_p/R_*=0.0786\pm0.0004$, and
$(M_{*}/M_{\odot})^{1/3}(R_*/R_{\odot})^{-1}=0.64\pm0.02$.  The
agreement between the two methods is thus good.

The times of the mid-points of the transits are given in
Table~\ref{tab:transit}. The errors in the measurements are on average
2.3 minutes, and the difference between the calculated times of the
transit are on average 2.9 minutes. There is no evidence of transit
timing variations.

\section{Excluding a background binary within the photometric mask}
\label{Sec:onoff}

Since the photometric mask of CoRoT has a size of
$30\arcsec\times16\arcsec$, we must confirm that the transits are 
caused by \object{CoRoT-19} and not by an eclipsing binary within the
photometric mask. To demonstrate this, we took images during transit
(``on'') and images out of transit (``off'') (Deeg et
al. \cite{deeg09}). If the transit were caused by an eclipsing binary
within the photometric mask, an eclipsing binary in the background
would be correspondingly fainter in the image taken during transit
than in the image taken out of transit. For this purpose, we used MONET
(located at McDonald Observatory, USA), the OGS a 1\,m telescope
(located at Iza\~na Tenerife, Spain), the 1\,m telescope of the
Florence George Wise Observatory (Israel), and the {\it Euler}
telescope (located at La Silla Chile).  We detected only two stars
within the mask, \object{CoRoT-19}, and a star that is labeled No. 5
in fig.~\ref{mask}. This star has $m_{R}=19.93\pm0.070$ mag, and leads
to a contamination factor of $\leq 0.32\%$. Star No.5 is thus to faint
to cause a transit with a depth of $0.79\pm0.02$\,\%.

After confirming that \object{CoRoT-19} was the object experiencing a
transit, we observed two additional transits with the TRAPPIST(Gillon et al.
\cite{gillon11}), and {\it Euler} telescopes to improve the accuracy of the
ephemeris. This step was important because \object{CoRoT-19b} was
observed by CoRoT for only 24 days.  The detection of the transit on
the target with ground-based telescopes confirms again that
\object{CoRoT-19} has a transiting planet orbiting with the period and
phase measured by CoRoT. The transit-times are given in
Table~\ref{tab:transit}.

\section{Spectroscopic observations}
\label{Sec:spectroscopy}

We obtained high-resolution echelle spectra of \object{CoRoT-19} 
to determine its mass, radius, and other host star properties, which 
were than used to derive the mass, radius and the equilibrium 
temperature of the planet, $T_{eq}$, of the planet. The
detection of the radial-velocity (RV) signal of the planet
confirms the nature of the photometric transit. All RVs obtained are
listed in Table~\ref{tab:RV}.  Fig.~\ref{RVmeas} shows the
phase-folded RV-measurements together with the orbit.

\subsection{Radial-velocity measurements}

\subsubsection {HARPS}

We took 19 spectra of \object{CoRoT-19} with the HARPS spectrograph
between 2010 December 15 and 2011 March 6  (Pepe et
al. \cite{pepe02}; Mayor et al. \cite{mayor03}) (ESO program
184.C-0639).  The spectra obtained have a resolving power
$\lambda/\Delta\lambda \approx 115\,000$ and cover the region from
3780\,\AA\, to 6910\,\AA\, in 72 spectral orders.  Eleven of the nineteen
HARPS spectra were taken during a transit in order to detect the
Rossiter-McLaughlin effect (Sect.~\ref{Sec:rossiter}), and the other spectra 
were used to determine the mass of the planet. The spectra were reduced 
and extracted using the HARPS pipeline (bias subtraction, flat-fielded,
scattered light subtraction, and wavelength calibration). The
sky-fibre was used to remove stray-light from the moon if
necessary. The RVs were determined by using a cross-correlation method
with a numerical mask that corresponds to a G2 star (Baranne et
al. \cite{baranne96}; Pepe et al. \cite{pepe02}).  The RV-measurements
were obtained by fitting a Gaussian function to the the average
cross-correlation function (CCF), after discarding the ten bluest and
the two reddest orders of the spectra that were of a low 
signal-to-noise ratio (S/N).

\subsubsection {SOPHIE}

We obtained seven spectra with the SOPHIE spectrograph on the 1.93\,m
telescope at the Observatoire de Haute-Provence, France (Bouchy et
al. \cite{bouchy09}; Perruchot \cite{perruchot08}). For our
observation, we used the high efficiency mode, which has a resolution of
$\lambda/\Delta\lambda=40\,000$ and covers the wavelength region from
3872\,\AA\, to 6943 \,\AA\, in 39 spectral orders (eight bluest and five reddest 
orders were discarded because of the low S/N). The reduction was
performed in a similar way to that for HARPS spectra.

\subsubsection {FIES}

Five spectra were also taken with the fibre-fed echelle spectrograph
FIES, which is mounted on the 2.6\,m Nordic Optical Telescope (NOT) at
Roque de los Muchachos Observatory (La Palma, Spain) in program
P42-216. We used the 1.3\arcsec \,high-resolution fibre, which gives a
resolving power of $\lambda/\Delta\lambda\approx67\,000$ and covers
the wavelength range 3600 - 7400~\AA. The data were reduced using
standard IRAF routines. The RVs were obtained using the RV standard
star HD\,50692 (Udry et al. \cite{udry99}), which was observed with the
same instrumental set-up.

\subsubsection {SANDIFORD}

Two additional RV measurements were performed on spectra
acquired with the Sandiford Cassegrain echelle spectrometer (SANDIFORD) 
mounted at the Cassegrain focus of the 2.1\,m (82 inch) Otto Struve Telescope 
at McDonald Observatory, Texas, USA. The spectra cover the wavelength range
5000-6000~\AA~with a resolving power of
$\lambda/\Delta\lambda\approx47\,000$. The adopted observing strategy,
data reduction, and RV-determination were the same as FIES.

\begin{figure}
  \includegraphics[height=.27\textheight,angle=0]{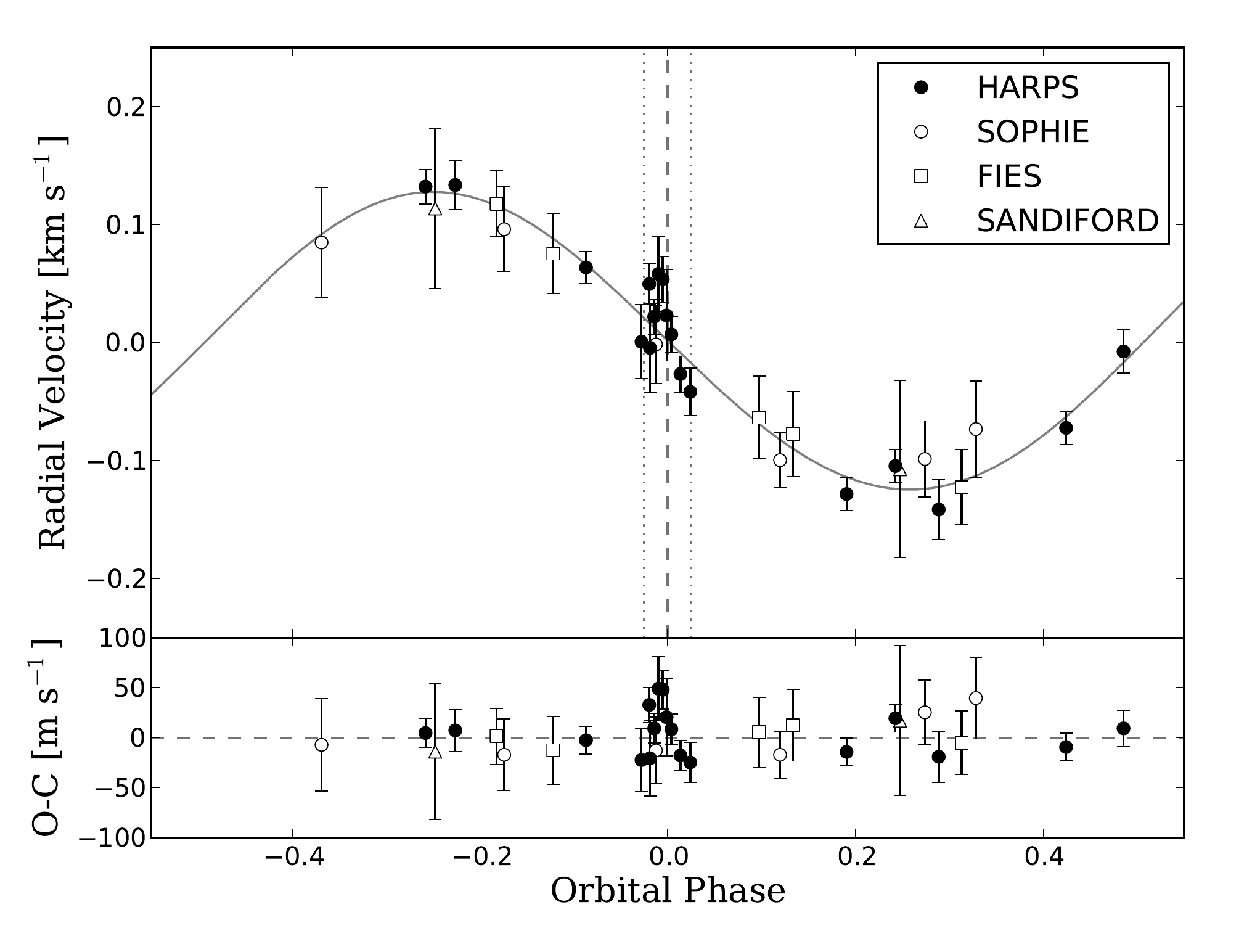}
\caption{RV-values together with the orbital fit (instrumental offset removed).}
  \label{RVmeas}
\end{figure} 

\begin{figure}
  \includegraphics[height=.27\textheight,angle=0]{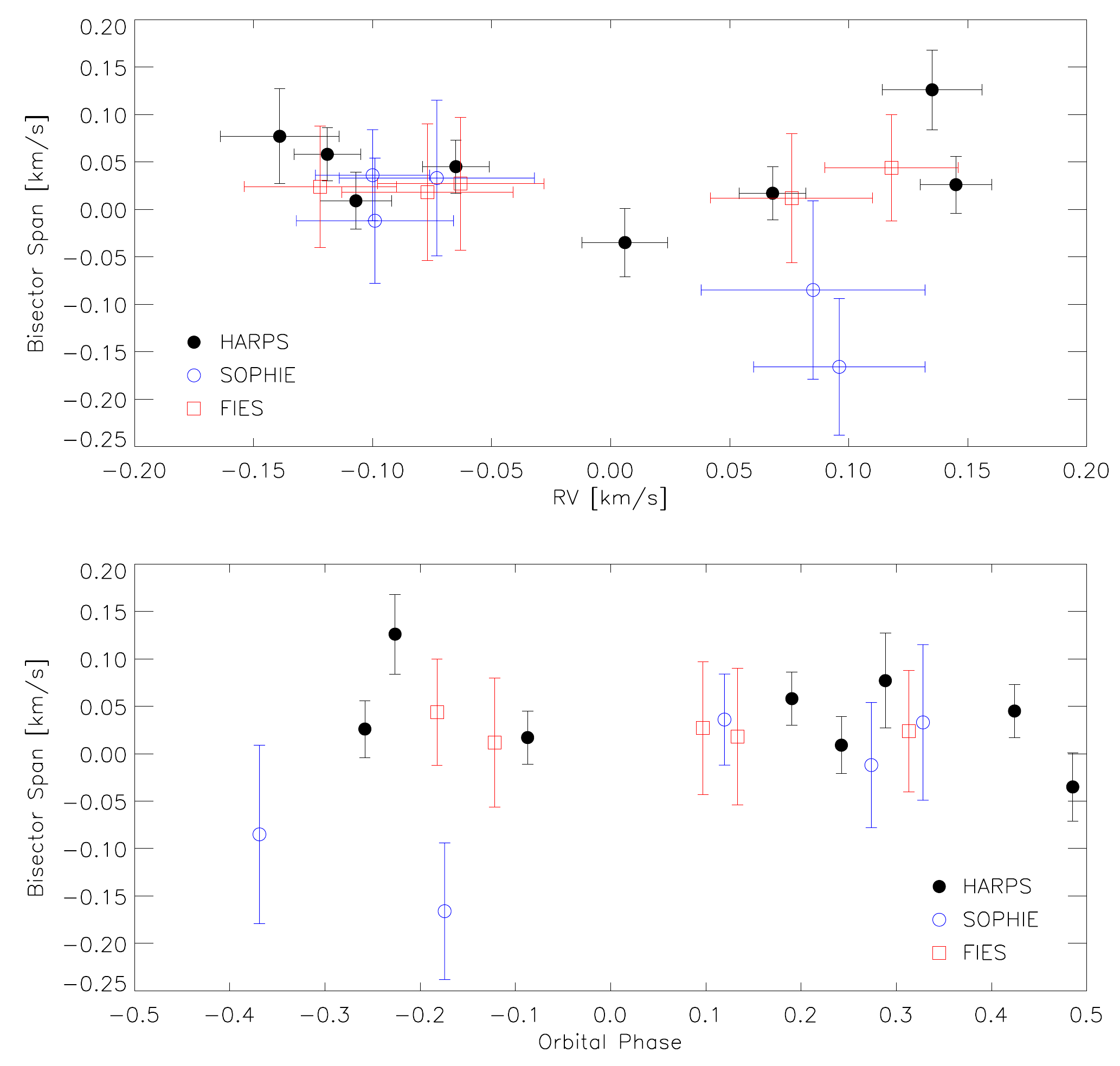}
\caption{Bisector span of the CCF versus the RV measurements (top panel) 
              and the orbital phase (bottom panel).}
  \label{RVBisctor}
\end{figure} 

\subsection{The mass of the planet and its orbit}
\label{Sec:RV}

In total, we obtained 33 spectra of \object{CoRoT-19} of which 12 were
taken during transit. For the orbit determination, we excluded
from the analysis all measurements taken during the transit, and the 
first SOPHIE measurement (BJD=2455513.6) because it was a clear
outlier.

Since the observations of the transits allow us to determine the
period and epoch to a very high accuracy, we used these values.  The
best-fitting orbit was derived by minimizing the $\chi^2$ using the
Downhill Simplex Algorithm (Nelder \& Mead \cite{nelder65}).  The
errors given in Table~\ref{tab:planet} correspond to the 68\%
confidence level using the bootstrap method.  The best-fit model curve
is shown in figure~\ref{RVmeas} and the parameters obtained are given
in Table~\ref {tab:planet}.

The scatter around the best-fit model orbit is comparable to the
typical error bar for each dataset. The errors were determined from
the photon noise and the wavelength calibration (Bouchy et
al. \cite{bouchy01}). The scatter around the best-fit solution is 12.2
$m\,s^{-1}$, 23.4 $m\,s^{-1}$, and 8.6 $m\, s^{-1}$ for HARPS, SOPHIE,
and FIES, respectively. The two SANDIFORD measurements differ on
average by 5.5 $m\,s^{-1}$ from the fitted curve (Table~\ref{tab:RV}),
with a confidence level of 95\% (resp. 99\%) that the orbital
eccentricity is smaller than 0.12 (0.15).

The detection of the RV-variations with the correct period and phase
lends strong support to the planet hypothesis. However, to
prove that the RV-variations are caused by the planet, we also
analyzed the spectral line asymmetry, which we measured by determining
the bisector of the CCF function of the HARPS spectra taken
out-of-transit (Queloz et al. \cite{queloz00}). The lack of
correlation between the bisector and the RV-variations shows that the
RV-variations are caused by the planet (Fig. \ref{RVBisctor}).

\begin{table}
\caption{Radial velocity measurements}
\begin{tabular}{l l l c  l}
\hline
\noalign{\smallskip}
HJD$^1$     & RV          & $\pm \sigma$ & BIS    & Instrument \\
-2 400 000  & [\kms]     & [\kms]             & [\kms] & \\
\hline
55546.7731 & -0.104 & 0.014 & -0.001 & HARPS$^2$ \\
55548.7204 & 0.133  & 0.015 & 0.050  & HARPS$^2$ \\
55574.7614 & -0.072 & 0.014 & 0.016  & HARPS$^2$ \\
55576.6662 & 0.064  & 0.014 & 0.005  & HARPS$^2$ \\
55577.7475 & -0.128 & 0.014 & 0.064  & HARPS$^2$ \\
55584.7230 & 0.050  & 0.018 & -0.011 & HARPS$^2$ \\
55586.6910 & -0.007 & 0.018 & -0.032 & HARPS$^2$ \\
55591.7124 & 0.134  & 0.021 & 0.065  & HARPS$^2$ \\
55592.5378 & 0.022  & 0.015 & 0.063  & HARPS$^2$ \\
55592.5735 & 0.054  & 0.019 & 0.006  & HARPS$^2$ \\
55592.6090 & 0.007  & 0.015 & 0.080  & HARPS$^2$ \\
55592.6467 & -0.026 & 0.015 & 0.151  & HARPS$^2$ \\
55592.6877 & -0.041 & 0.020 & 0.007  & HARPS$^2$ \\
55593.7185 & -0.141 & 0.025 & 0.083  & HARPS$^2$ \\
55627.5593 & 0.001  & 0.031 & -0.088 & HARPS$^2$ \\
55627.5944 & -0.004 & 0.037 & -0.003 & HARPS$^2$ \\
55627.6298 & 0.059  & 0.032 & 0.048  & HARPS$^2$ \\
55627.6633 & 0.023  & 0.039 & 0.066  & HARPS$^2$ \\
\hline
55513.6044 &  0.239 & 0.028 &  0.012  & SOPHIE$^3$ \\
55519.6156 & -0.099 & 0.033 & -0.012  & SOPHIE$^3$ \\
55527.6213 & -0.073 & 0.041 &  0.033 & SOPHIE$^3$ \\
55529.5611 &  0.096 & 0.036 & -0.166  & SOPHIE$^3$ \\
55557.4709 & -0.002 & 0.034 & -0.071  & SOPHIE$^3$ \\
55577.4717 & -0.100 & 0.024 &  0.036  & SOPHIE$^3$ \\
55579.4660 &  0.085 & 0.047 & -0.085  & SOPHIE$^3$ \\
\hline
55568.5012 &  0.118 & 0.028 & 0.044 & FIES$^4$ \\
55569.5898 & -0.063 & 0.035 & 0.027 & FIES$^4$ \\
55570.4314 & -0.122 & 0.032 & 0.024 & FIES$^4$ \\
55580.4278 &  0.076 & 0.034 & 0.012 & FIES$^4$ \\
55581.4217 & -0.077 & 0.036 & 0.018 & FIES$^4$ \\
\hline
55585.7638 & -0.107 & 0.075 & -- & SANDIFORD$^5$ \\
55587.7321 & 0.114 & 0.068 & -- & SANDIFORD$^5$ \\
\hline
\end{tabular}
\label{tab:RV}
\\
$^1$ The Heliocentric Julian date is calculated directly from the UTC.\\
$^2$ HARPS: $V_0 = 24.228 \pm0.002$ \kms \\
$^3$ SOPHIE: $V_0 = 24.192 \pm 0.002$ \kms \\
$^4$ FIES: $V_0 = 24.065 \pm0.005$ \kms \\
$^5$ SANDIFORD: $V_0 = 24.324 \pm0.003$ \kms \\
\end{table}

\begin{table}
 \centering
 \caption{Abundances of some chemical elements for the fitted lines in the
  \emph{HARPS} co-added spectrum. The abundances refer to the solar value 
  and the last column reports the number of lines used.
 \label{tab:abund}}
 \setlength{\tabcolsep}{3pt} 
\begin{tabular}{lcr}
\hline
\noalign{\smallskip}
       Element & Abundance & No. lines \\
\noalign{\smallskip}
\hline
\noalign{\smallskip}
  {Ca \sc   i} &  $  0.08   \pm 0.11 $ &   6  \\ 
  {Y  \sc  ii} &  $ -0.03   \pm 0.15 $ &   3  \\ 
  {Sc \sc  ii} &  $  0.00   \pm 0.11 $ &   4  \\ 
  {Ti \sc   i} &  $  0.00   \pm 0.11 $ &  10  \\ 
  {Ti \sc  ii} &  $  0.08   \pm 0.11 $ &   4  \\ 
  {Cr \sc   i} &  $ -0.03   \pm 0.12 $ &   8  \\ 
  {Fe \sc   i} &  $ -0.02   \pm 0.10 $ & 134  \\ 
  {Fe \sc  ii} &  $ -0.02   \pm 0.11 $ &  13  \\ 
  {Ni \sc   i} &  $ -0.04   \pm 0.10 $ &  18  \\ 
  {Si \sc   i} &  $ -0.02   \pm 0.10 $ &  10  \\ 
\noalign{\smallskip}
\hline
\end{tabular}
\end{table}

\begin{table}
 \caption{Parameters of the planet and the star}
\setlength{\tabcolsep}{3pt} 
 \begin{tabular}{l l}
 \hline 
 \hline 
Ephemeris & \\
 \hline 
Orbital period P [days] & $3.89713\pm0.00002$ \\                                  
Epoch $T_{0}$ [HJD] & $2~455~ 257.44102\pm0.0006^{(1)}$ \\ 
Transit duration $d_{tr}$ [h] & $4.7\pm0.1$ \\ 
& \\
{\em Derived parameters from} & \\
{\em radial velocity observations} & \\
Orbital eccentricity e & $0.047\pm0.045$ \\
RV semi-amplitude K [$ms^{−1}$] & $126\pm6$ \\
Systemic velocity $V_{R}$ [$kms^{−1}$] & $24.16\pm0.09$ \\
O−C residuals [$ms^{−1}$] & 15 \\
Semi-major axis a [AU] & $0.0518\pm0.0008$ \\
& \\
{\em Transit parameters}      & \\
$k=R_p/R_*$                          & $0.0786\pm0.0004^{(2)}$\\
Scaled semi-major axis $a/R_*$ & $6.7\pm0.1^{(2)}$ \\
$(M_{*}/M_{\odot})^{1/3} (R_*/R_{\odot})^{-1}$ & $0.64\pm0.02^{(2)}$ \\
Impact parameter $b^{(4)}$ & $0.24\pm0.08^{(2)}$ \\
Inclination $i$ [deg] & $88.0\pm0.7$\\
Linear limb darkening coefficient $u_{+}$ & $0.58\pm0.02$ \\
Linear limb darkening coefficient $u_{-}$ & $0.09\pm0.04$ \\
& \\
{\em Spectroscopic parameters of the star} & \\
Effective temperature $T_{\rm eff}$ [K]  & $6090\pm70$\\
Surface gravity $\log\,g$ [cgs] & $4.07\pm0.03^{(3)}$ \\
Metallicity ${\rm [Fe/H]}$ [dex]  & $-0.02\pm0.10$ \\
$v\,sin\,i_*$ [$km\,s^{−1}$]  & $6\pm1$ \\
$v_{micro}$ [$km\,s^{−1}$] & $1\pm1$\\
$v_{macro}$ [$km\,s^{−1}$] & $4\pm1$\\
Spectral type & F9V \\
& \\
{\em Physical parameters} & \\
$M_v$ & $3.40\pm0.05^{(3)}$ \\
$M_{bol}$ & $3.38\pm0.05$ \\
$L/L_{\odot}$ & $3.5\pm0.2$ \\
B-V   & $0.56\pm0.02$ \\
interstellar extinction AV [mag] & $1.2\pm0.5$\\
Distance of the star d [pc] & $770\pm160$ \\
Star mass [$M_\odot$] &  $1.21\pm0.05^{(3)}$ \\
Star radius [$R_\odot$] & $1.65\pm0.04^{(3)}$ \\
Mean stellar density $\rho_*$ [$g\,cm^{-3}$]  & $0.38\pm0.03$ \\
Age of the star [Gyr] & $5\pm1^{(3)}$ \\
Planet mass $M_p$ [$M_{Jup}$] & $1.11\pm0.06$ \\
Planet radius $R_p$ [$R_{Jup}$] & $1.29\pm0.03$ \\
Planet density $\rho_{p}$ [$g\,cm^{-3}$] & $0.71\pm0.06$ \\
Equilibrium temperature at & \\
the averaged distance $T_{eq}$ [K] & $2000\pm150$ \\
$\lambda^{(5)}$  [deg] & $-52^{+27}_{-22}$ \\
 \hline 
 \end{tabular}
 \label{tab:planet}
\\
$^1$ Heliocentric Julian date \\
$^2$ Derived from light-curve fitting using the MCMC and genetic algorithm \\
$^3$ Derived by using the evolutionary tracks published by Girardi et al. (\cite{girardi00}). 
     The values are agreement with STAREVOL tracks \\
$^4$ Impact parameter is defined as $b={\frac{a\cdot cos\,i}{R_*}}$\\
$^5$ sky-projected angle between the planetary orbital axis and the stellar rotation axis \\
\\
\end{table}

\subsection{The properties of the host star}
\label{Sec:SpectralAnalysis}

To determine the photospheric parameters of
\object{CoRoT-19}, we constructed a master spectrum by co-adding the
HARPS spectra after shifting them to the same RV zero point.  The
master spectrum has a S/N of 130 at 5500 \AA .
  
Following the method described in Deleuil et al.  (\cite{deleuil08})
and Bruntt et al. (\cite{bruntt10}), the spectral analysis was
performed using the semi-automatic package VWA (Bruntt et al.
\cite{bruntt02}; \cite{bruntt08}; \cite{bruntt10}). The parameters of
the star are $T_{\rm eff}=6090\pm70$\,K, $\log\,g=4.0\pm0.1$ [cgs],
and $[Fe/H]= -0.02\pm0.10$ [dex]. Using the {\it Spectroscopy Made
  Easy} (SME) method (Valenti \& Piskunov \cite{valenti96}), we
obtained $T_{\rm eff}=6200\pm70$\,K, $\log\,g=4.2\pm0.1$ [cgs], and
$[M/H]= 0.1\pm0.1$ [dex], which is in agreement with the VWA method.
The abundances derived are given in Table~\ref{tab:abund}, and all other
parameters of the star are given in Table~\ref{tab:planet}).

Instead of using $\log\,g$ from the spectroscopic analysis to
determine the mass and radius of the star, we used the value
$M^{1/3}R^{-1}$ obtained from the light-curve analysis, 
and the spectroscopically derived $T_{\rm eff}$, and ${\rm [Fe/H]}$  
because the log(g) determined in this way has a smaller error than the
spectroscopic determination. Since the mass and radius of the star
are derived from the evolutionary tracks, the parameters of the
planets also depend on the accuracy of the tracks. We thus determined
these parameters using two different sets of tracks. The STAREVOL
evolution tracks (Palacios, priv. comm.) yield a mass of
$M_*=1.20\pm0.06\,M_\odot$, and a radius of 
$R_*=1.62\pm0.08\,R_\odot$ for the star.

We also used a Bayesian estimation of stellar parameters (da Silva et
al.  \cite{silva06}) and the evolutionary tracks published by Girardi
et al. (\cite{girardi00}) to determine the stellar
parameters. These tracks were tested by Baines et
al. (\cite{baines10}) by comparing the diameters of stars derived from
the tracks with the diameters measured directly with the CHARA
interferometer. Using these tracks, we obtained
$M_*=1.21\pm0.05\,M_\odot$, and $R_*=1.65\pm0.04\,R_\odot$, 
which are values entirely consistent with those from STAREVOL.

By combining the radius of the star with the relative size of the
planet determined from the transit curve (Table~\ref {tab:planet}), we
derived a radius for the planet of $R_{p}=1.29\pm0.03\,R_{Jup}$.  The
absolute magnitude of the star is $M_v=3.40\pm0.05$, and the distance
$770\pm160\,pc$, if we take the extinction of $Av=1.2\pm0.5$ into
account.  The star is close to the end of the H-burning phase and has
an age from four to six Gyrs.  Using the HARPS spectrum, we obtained
$v\,sin\,i=6\pm1\,km\,s^{-1}$. This means that the rotation period of
the star is $P_{rot}\geq14\,d$. The star was not detected in the ROSAT
all-sky survey and we did not detect either the Li\,{\sc I} line at
6707.8\,\AA\ or an emission component in the Ca\,{\sc II}\,H \&K lines
in the master spectrum.  In summary, the host star is an old,
inactive F9\,V star close to the end of its life on the main-
sequence.

\section{The Rossiter McLaughlin effect}
\label{Sec:rossiter}

\begin{figure}
  \includegraphics[height=.25\textheight,angle=0]{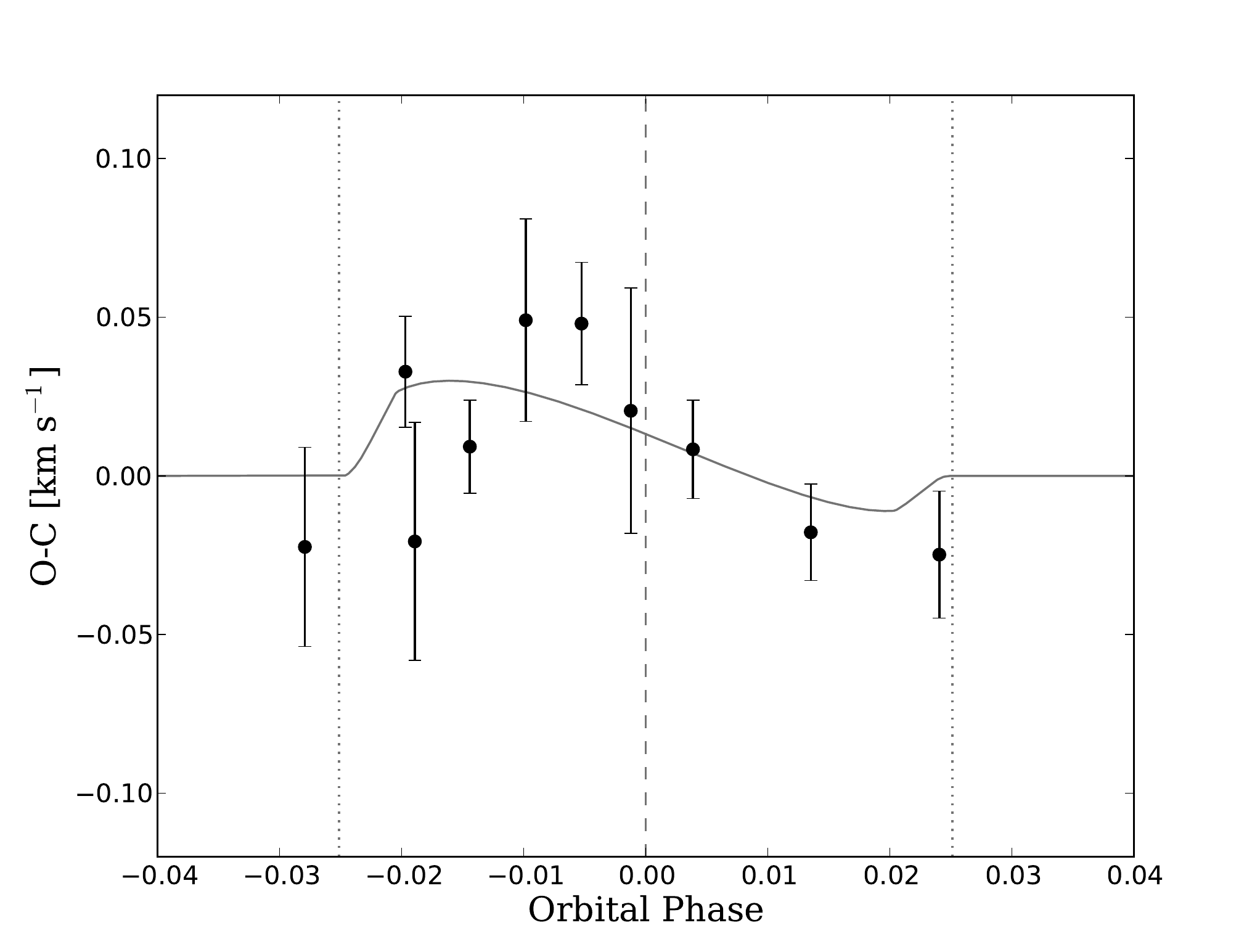}
\caption{RV-measurements taken during transit with the orbit subtracted
         (Rossiter-McLaughlin effect).}
  \label{rossiter}
\end{figure}

The spectroscopic transit \object{CoRoT-19} was observed with HARPS 
to detect the Rossiter-McLaughlin (RM) anomaly. This apparent
distortion of the stellar lines profile owing to the transit of a
planet in front of a rotating star allows us to measure the
sky-projected angle between the planetary orbital axis and the stellar
rotation axis $\lambda$ (e.g. Bouchy et al. \cite{bouchy08}).
Fig.~\ref{rossiter} shows the RV-measurements obtained with HARPS
after subtracting the orbit. The uncertainties were determined from
the obtained $\chi^2$ surface, taking into account the errors in the
K-amplitude, $V_r$, and $i$.  The RM effect is marginally detected (at
$2.3 \sigma$) with $\lambda=-52^{+27}_{-22}$ deg.  There is thus a
hint that the system is misaligned but to only at the $2\sigma$-level.

\section{Discussion and conclusions}

 \begin{figure}
  \includegraphics[height=.27\textheight,angle=0]{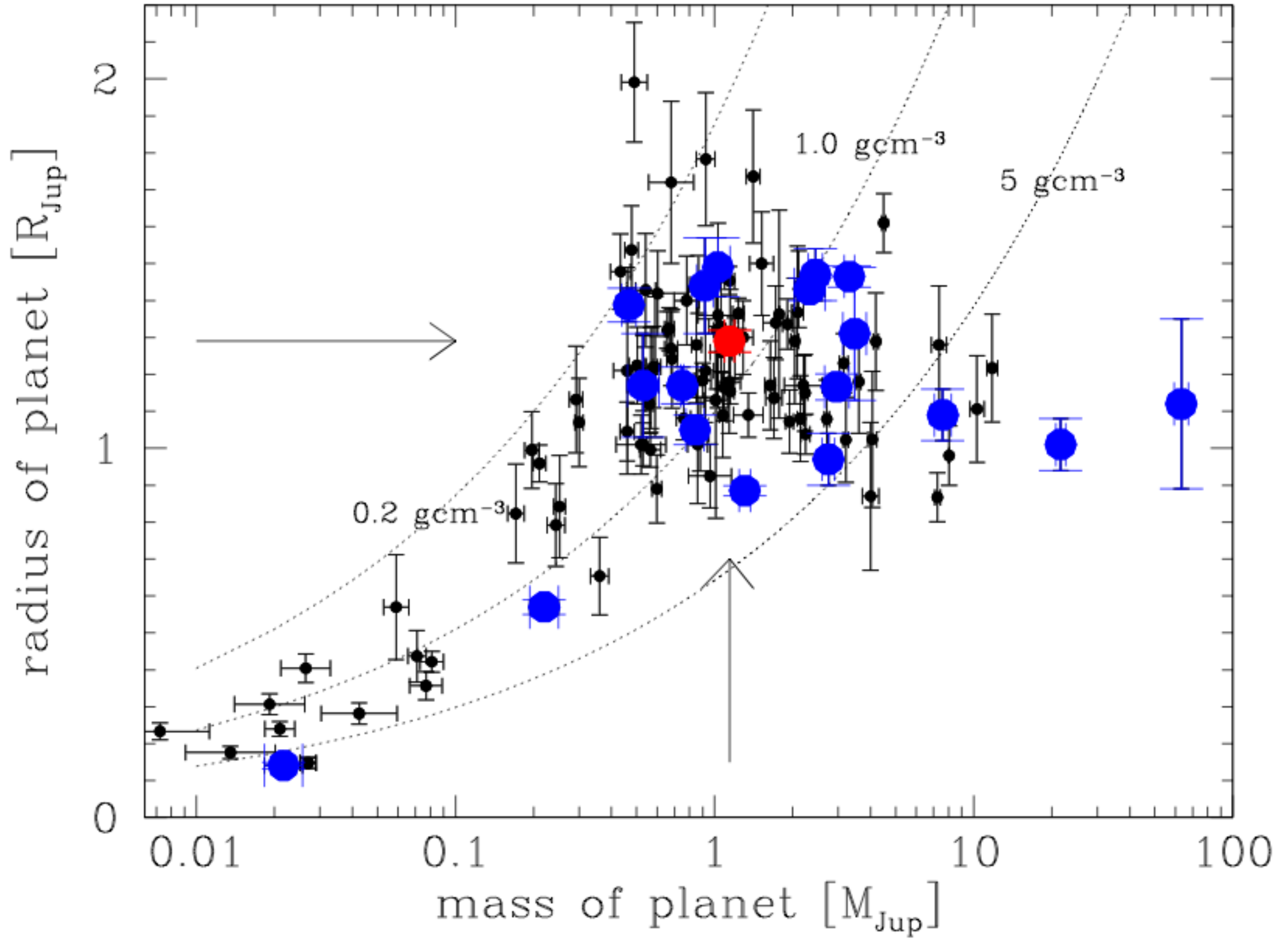}
 \caption{Mass-radius diagram of the planets and brown dwarfs. 
              CoRoT-19b is marked with a thick red point and by the arrows, all other
              objects discovered by CoRoT are indicated by the tick blue points, and
              planets discovered in other surveys are shown as 
              small (black) points.
              }
  \label{Aplanets}
\end{figure}

In Fig.~\ref{Aplanets} we present the mass-radius diagram for transiting
planets and brown dwarfs.  The objects marked with a large blue point
are those discovered by CoRoT, the objects with small points are those
detected with other instruments. The planet \object{CoRoT-19b} is 
indicted by a large red dot and the two arrows. The two CoRoT-planets above 
\object{CoRoT-19b} are \object{CoRoT-1b} (Barge et al. \cite{barge08})
and \object{CoRoT-12b} Gillon et al. (\cite{gillon10}). Their
densities are $0.38\pm0.05$ and $0.31\pm0.08$ $g\,cm^{-1}$,
respectively, a bit lower than \object{CoRoT-19b}, which has
$0.71\pm0.06$ $g\,cm^{-1}$. Thus, \object{CoRoT-19b} is an example 
of an inflated extrasolar planet but is less extreme than the others.

\begin{acknowledgements}

The French teams are grateful to CNES for its constant support and
funding of AB, JMA, and CC.  The German team thanks the DLR and the
BMBF for the support under grants 50 OW 0204, and 50 OW 0603. The team
at the IAC acknowledges support by grants ESP2007-65480-C02-02 and
AYA2010-20982-C02-02 of the Spanish Ministerio de Ciencia e
Innovaci\'on.  TRAPPIST is funded by the Belgian Fund for Scientific
Research (Fond National de la Recherche Scientifique, FNRS) under the
grant FRFC 2.5.594.09.F, with the participation of the Swiss National
Science Foundation (SNF). M. Gillon and E. Jehin are FNRS Research
Associates. MONET (MOnitoring NEtwork of Telescopes) is funded by the
"Astronomie \& Internet" program of the Alfred Krupp von Bohlen und
Halbach Foundation, Essen, and operated by the
Georg-August-Universit\"at G\"ottingen, the McDonald Observatory of
the University of Texas at Austin, and the South African Astronomical
Observatory. The results are also partly based on observations
performed with the FIES spectrograph at the Nordic Optical Telescope
(NOT), under observing program P42-261. NOT is operated on the island
of La Palma jointly by Denmark, Finland, Iceland, Norway, and Sweden,
in the Spanish Observatorio del Roque de los Muchachos of the
Instituto de Astrofisica de Canarias. Additional data was obtained
with the Sandiford spectrograph at the 2.1\,m Otto Struve telescope at
McDonald Observatory (Texas, USA), and the SOPHIE spectrograph at the
Observatoire de Haute-Provence, France, under observing program
PNP.08A.MOUT.

\end{acknowledgements}

\end{document}